# Multi-Objective Optimization for Common-Centroid Placement of Analog Transistors

Supriyo Maji
The University of Texas at Austin
Austin, USA
smaji@alumni.purdue.edu

Hyungjoo Park
Hanyang University
Seoul, South Korea
pikkoro97@hanyang.ac.kr

Gi moon Hong
The University of Texas at Austin
Austin, USA
gimoon.hong@austin.utexas.edu

Souradip Poddar
The University of Texas at Austin
Austin, USA
souradippddr1@utexas.edu

David Z. Pan
The University of Texas at Austin
Austin, USA
dpan@ece.utexas.edu

## ABSTRACT
In analog circuits, process variation can cause unpredictability in circuit performance. Common-centroid (CC) type layouts have been shown to mitigate process-induced variations and are widely used to match circuit elements. Nevertheless, selecting the most suitable CC topology necessitates careful consideration of important layout constraints. Manual handling of these constraints becomes challenging, especially with large size problems. State-of-the-art CC placement methods lack an optimization framework to handle important layout constraints collectively. They also require manual efforts and consequently, the solutions can be suboptimal. To address this, we propose a unified framework based on multi-objective optimization for CC placement of analog transistors. Our method handles various constraints, including degree of dispersion, routing complexity, diffusion sharing, and layout dependent effects. The multi-objective optimization provides better handling of the objectives when compared to single-objective optimization. Moreover, compared to existing methods, our method explores more CC topologies. Post-layout simulation results show better performance compared to state-of-the-art techniques in generating CC layouts.

## 1 INTRODUCTION

In analog circuits, process variation can affect matching of the devices, which degrades circuit performance [10]. Process-induced variations can be categorized as random variations and systematic variations. One effective technique to reduce random variations is to make the devices bigger [10][24]. However, increasing the size may cause the devices to become more sensitive to process gradients [24]. CC layouts have been shown to be better than other alternatives such as clustered and interdigitated patterns for mitigating process-induced linear variations [26] and are widely used to match circuit elements [24][30][7][16][15]. In a CC layout, each device is broken into multiple units and the units are placed in different locations in an array in such a way that the spatial centroids of the devices overlap. A CC layout is symmetric about both the X and Y-axes.

However, the challenge in implementing a CC type layout comes from the fact that there can be many CC topology configurations, i.e., the devices can be placed in many different ways for the centroids to overlap. Moreover, one must consider various layout constraints while optimizing a CC layout. These include maximizing degree of dispersion to achieve uniform device spread, a factor affecting variation performance [24], minimizing route length to reduce parasitics and voltage drop, maximizing diffusion sharing to reduce layout area, and minimizing layout dependent effects such as Length of Diffusion (LOD) and Well Proximity Effects (WPE) to mitigate threshold voltage change. Considering these competing constraints, manually selecting the optimal CC topology becomes challenging [21][11].

Over the last decade, various studies have delved into CC placement of analog devices including transistors, capacitors and resistors. Some earlier approaches have focused on generating high-quality CC topologies, mainly considering spatial variation, not layout effects [8][24][28][3][19]. On the other hand, methods proposed in [31][17][28][4][30][13][12] may not apply to general transistor circuits. The recent works presented in [26][25][11] have made substantial progress in incorporating layout constraints into CC placement. However, these constraints are addressed separately through post-processing steps following CC topology generation. For instance, to enhance circuit offset performance, dummy components are placed around the CC structure, effects of parasitics and electromigration are taken into account during routing phase following the placement step [22][25]. One significant drawback is the lack of an optimization framework capable of collectively addressing the constraints, which can lead to suboptimal results. The work in [22] introduces a simulated annealing-based optimization framework that handles several layout constraints while optimizing nonlinear spatial variation. However, the final layout type achieved using this method is not CC. Unlike [26][25][11], where the fixed nature of the formulation limits exploring CC topologies, the simulated annealing-based approach in [22] allows exploration of various non-CC topologies. While [26][25][11] present post-layout simulation results, the experimental findings in [22] are model-based.

We introduce a unified multi-objective optimization framework to generate CC-type layout of analog transistors while handling important layout constraints. Unlike single-objective optimization method, which requires careful tuning of coefficients to balance different objectives, multi-objective approach eliminates the need for tuning. Specifically, we enhance a well-known multi-objective optimization algorithm AMOSA [1]. AMOSA has been used for solving circuit-level placement problem [23]. Here, we use it to address device-level placement problem. AMOSA relies on the concept of the amount of domination rather than coefficient-based control of objectives. The use of an archive to store non-dominating solutions seen during the optimization process allows diverse exploration of solution space. Moreover, compared to [26][25], we explore more CC topologies by applying powerful transformations.

The key contributions of our work are as follows:

- To the best of our knowledge, we are the first to propose a unified multi-objective optimization framework for CC placement of analog transistors.
- We explore more CC topologies compared to the state-of-the-art by applying powerful transformations.
- Our optimization formulation encompasses important layout constraints, including diffusion break, layout dependent effects, routing cost, and degree of dispersion.
- Post-layout simulation results show that the proposed method performs better than state-of-the-art across different circuit configurations and important constraints.

The rest of the paper is organized as follows. Section 2 discusses our CC placement optimization framework. Section 3 presents various layout constraints handled by our framework. Results are discussed in Section 4. Section 5 concludes the paper.

## 2 CC PLACEMENT OPTIMIZATION

Simulated annealing, a classical optimization technique [14], has been used for both digital [6][29] and analog circuit placement [31][18][20][22]. The main idea is to optimize circuit performance by perturbing potential placement solutions with actions such as random selection, swapping, and rotation. However, the single objective optimization technique in these approaches hinges on a cost function. The selection of coefficient values for various objectives in the cost function is a manual task. In [1], an algorithm AMOSA for multi-objective optimization based on simulated annealing has been proposed. AMOSA uses a concept of amount of domination to compute the acceptance probability of a new solution. It utilizes an archive to retain the non-dominated solutions encountered so far. To better understand the concept, consider two solutions, denoted as $sol_1$ and $sol_2$. $sol_1$ dominates $sol_2$, if $\forall i \in \{1, 2, ..., M\}$, $f_i(sol_1) \leq f_i(sol_2)$, where $f$ is the objective to be minimized and $M$ is the number of objectives. For the two solutions, the amount of domination is defined as follows.

$$\Delta dom_{sol_1, sol_2} = \prod_{i=1, f_i(sol_1) \neq f_i(sol_2)}^{M} \frac{|f_i(sol_1) - f_i(sol_2)|}{R_i} \quad (1)$$

Here, $R_i$ represents the range of the $i^{th}$ objective. The algorithm begins by entering an initial solution, termed cur-pt, into the archive at temperature $T_{max}$. The cur-pt is perturbed to yield a new solution, referred to as new-pt. The domination status of the new-pt is then checked with respect to the cur-pt and the solutions within the archive. The archive and the cur-pt are updated based on the domination status. This process iterates a total of $n$ times for each temperature. The temperature is reduced to $\alpha$ x $temp$, using the cooling rate $\alpha$, until the minimum temperature, $T_{min}$, is reached. Upon reaching $T_{min}$, the iteration concludes, and the archive contains the final non-dominated solutions. Post-processing can be applied to this archive to get the most desired solution.

### 2.1 Initial Placement with Min. Diffusion Break

Diffusion sharing is a widely used concept in analog applications [25]. Sharing diffusion region not only reduces layout area and routing cost but also minimizes spatial variation [22]. However, fully sharing diffusion region without using dummies is not always possible in a CC type layout, as diffusion breaks could be unavoidable due to other constraints [25][26][30][22]. We present two cases in Figure 1(a): one where diffusion break can be avoided and the other where it is likely to occur, necessitating the use of a dummy transistor. A lower value of diffusion break does not always translate to a smaller overall area. The location of the diffusion break can impact the need for additional dummy transistors to maintain a CC structure while sharing diffusion region. For example, shown in Figure 1(b), compared to solution 2, solution 1 requires more layout area and routing resources, although it has fewer diffusion breaks.

However, our findings indicate that minimizing only the dummy count leads to inferior results. We, therefore, optimize both the dummy count and the diffusion break. First, we generate an initial placement that has a minimum number of diffusion breaks. Subsequently, during optimization, a new solution is accepted only if the diffusion break and the number of dummies do not increase beyond an upper bound, which can be a user-defined constraint. We create the initial CC placement (i.e. the cur-pt) by placing half of the devices and mirroring the other half (XX/180° transformation, where X is a device [24]). For placing half the devices, we adopt a sequential placement approach, grouping units of the same device together. Consecutive devices share drain or source terminal. This is illustrated in Figure 2. To generate the new solution (i.e. the new-pt), we randomly swap two distinct unit devices within half of the device set and mirror this action in the other half.

### 2.2 CC Topology Space Exploration

However, the mirroring strategy of device placement using the XX/180° transformation proves insufficient. To explore more CC topologies, we also employ XY/$180°$ transformation, where X and Y are devices [24]. The results of these two transformations for a two-transistor layout with four units each yield six distinct layout types, four of which adhere to the common-centroid configuration, as shown in Figure 3. We have not shown the XX/$180°$ or XY/$180°$ transformation on "B A A B" or "A B A B" as they would not produce any new topology or distinct pattern. Due to limitations in the algorithm, the methodology proposed in [25] can only produce one CC topology. While there are many other transformations possible through rotational and reflectional symmetries, the XX/$180°$ or XY/$180°$ transformations are considered the most powerful [24]. These transformations are applied at the perturbation stage in each iteration of the multi-objective optimization algorithm run. The handling of transformation in a general case of perturbation that includes random swap is illustrated in Figure 4. Note that XY/$180°$ transformation can be applied only to devices having the same number of units. If a transformation yields a non-CC type layout, it is not considered a solution.

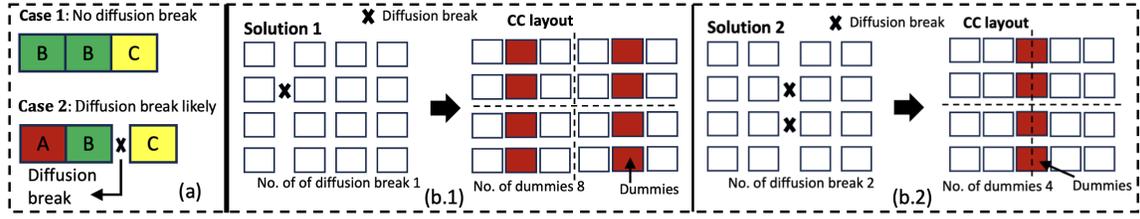

**Figure 1:** (a) Case 1: No diffusion break, Case 2: Diffusion break likely. Here A, B and C are unit transistors sharing drain or source terminal. (b.1 & b.2) Solution 1 has one diffusion break compared to two in solution 2, however, solution 1 requires eight dummy insertions compared to four in solution 2 to maintain CC structure while sharing diffusion region.

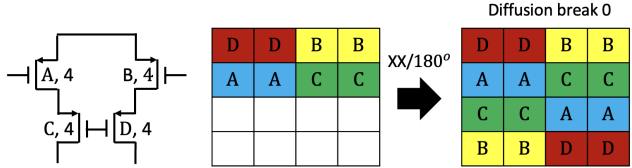

**Figure 2:** Generating initial CC placement that has minimum number of diffusion break for the circuit on the left by applying XX/$180°$ (where, X ∈ {A, B, C, D}) transformation on half of the devices placed sequentially.

### 2.3 Enhancement to AMOSA

We enhance the original AMOSA algorithm [1] to handle a specific situation in our problem where the perturbation stage may generate more than one new solution. Note that the XX/$180°$ transformation yields only one solution, while the XY/$180°$ transformation can generate multiple solutions as each pair of devices with the same number of units can produce a unique topology. For instance, in Figure 4, each combination <A B>, <A C>, <A D>, <B C>, <B D>, <C D> produces a distinct topology, although only <B C> qualifies as CC. Unlike AMOSA, which generates one new solution in each iteration and stores it in new-pt, we store the solutions in new-pts. Subsequently, we have the task of updating the cur-pt and the archive. We consider three cases for updating the cur-pt. **Case 1:** cur-pt dominates $k_1$ (> 0) solutions in new-pts, **Case 2:** cur-pt non-dominates all solutions in new-pts and **Case 3:** $k_1$ (> 0) solutions in new-pts dominate cur-pt. At any iteration, new-pts and archive contain only non-dominated solutions.

- **Case 1:** Given that $k_1$ solutions out of a total $k$ solutions in new-pts are dominated by cur-pt, the remaining $k - k_1$ solutions are, by transitivity, non-dominating. Therefore, we randomly choose one solution from $k - k_1$ solutions and assign it to cur-pt with some probability. The probability ($prob$) calculation considers the degree of domination of $k_1$ solutions in new-pts by cur-pt and the solutions in the archive. We modify Eq. (2) in [1] as follows.

$$prob = \frac{1}{1 + exp(\frac{\Delta dom_{avg}}{temp})} \quad (2)$$

Where,

$$\Delta dom_{avg} = \frac{(\sum_{i=1}^{k_3} \sum_{j=1}^{k_2} \Delta dom_{i,j}) + \sum_{i=1}^{k_1} \Delta dom_{\text{cur--pt},i}}{k_4 + k_1}$$

Here, $k_2$ is the number of solutions in the archive dominating some solutions in new-pts, $k_3$ is the number of solutions in new-pts dominated by some solutions in the archive, and $k_4$ is the number of such domination. Note that probability calculation considers more exploration around new solution at higher temperature, which is typical for a simulated annealing algorithm [14].

- **Case 2:** Since all solutions in new-pts are non-dominating w.r.t. cur-pt, we randomly pick one solution and assign it to cur-pt based on the probability in Eq. (2) with $k_1 = 0$.
- **Case 3:** Since there are solutions in new-pts dominating cur-pt, one of them is chosen randomly and assigned to cur-pt.

Next, for updating the archive, we use solutions in new-pts to replace dominated solutions in the archive. Additionally, while the original algorithm incorporates clustering to mitigate the loss of diversity in solutions, we have chosen not to employ clustering due to the relatively small size of the solution set in our problem. A pseudocode for our CC placement optimization algorithm is presented in **Algorithm 1**.

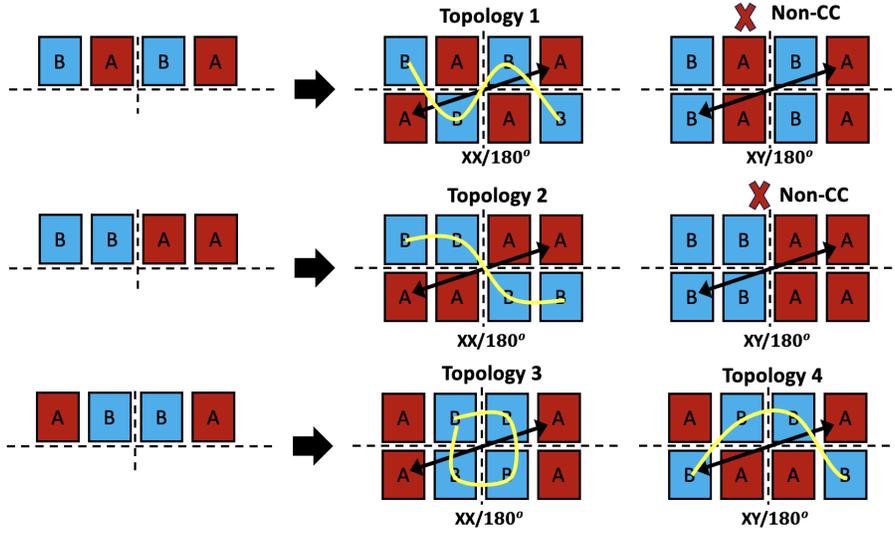

**Figure 3:** Applying **XX**/180° and **XY**/180° (where, X, Y ∈ {A, B, C, D}, X ≠ Y) transformations on a two transistors schematic with four units each. Out of the six layouts produced by the transformation four are CC type.

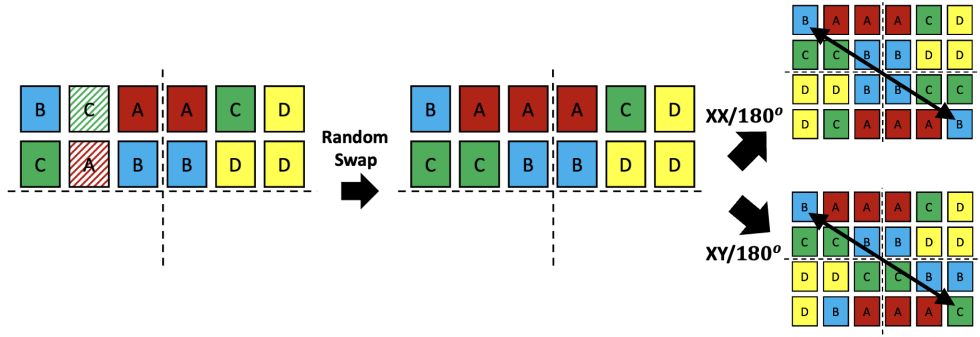

**Figure 4:** Random swap (A <=> C) and **XX**/180° and **XY**/180° transformations for a more general case of perturbation. XY/180° transformation yields CC topology only when applied on devices B and C.

## 3 LAYOUT CONSTRAINTS

Next we discuss the different layout constraints, i.e. the objectives functions ($f_i$ for $i = 1$ to $M$ in Eq. (1)) to be handled by the proposed optimization algorithm.

### 3.1 Degree of Dispersion

One of the fundamental rules for CC layouts underscores the importance of achieving the maximum degree of dispersion [10]. This involves distributing the device uniformly throughout the array. In [24], a quantitative measure of the degree of dispersion has been proposed. Assume there is an edge between adjacent units in a placement solution. Then, the following expression captures the degree of dispersion.

$$\frac{2 \sum OK - 2n_c n_r + (n_c + n_r)}{2n_c n_r - (n_c + n_r)} \quad (3)$$

Here, if an edge connects units belonging to different devices, OK is 1, otherwise OK is 0. $n_c$ and $n_r$ are respectively the number of columns and rows in a CC placement. This is illustrated in Figure 5. The Degree of Dispersion, measured within the range (-1, 1], signifies a higher value as being better. A higher value indicates increased spreading of the device units throughout the layout, resulting in reduced clustering. In Figure 5, in the Table we show the degree of dispersion values for the four topologies. Among the four, topology 1 has the highest dispersion value, reflecting a uniform spread of both transistors along both the X and Y axes. Topology 3 exhibits the lowest dispersion, attributed to the clustering of B units along both the X and Y axes. Comparing topologies 4 and 2, the latter exhibits more clustering along the X axis. When comparing topologies 1 and 4, the former exhibits a better interdigitation along the X-axis.

### 3.2 Layout Dependent Effects (LDE)

We account for the Well Proximity Effect (WPE) as it is a critical Layout Dependent Effects (LDE) [27][9]. WPE captures variations in the threshold voltage based on the distance of

## Algorithm 1 CC Placement

1: Set $T_{max}$, $T_{min}$, iter, $\alpha$, temp = $T_{max}$
2: Get cur-pt with minimum diffusion break and add it to the archive
3: **while** temp > $T_{min}$ **do**
4:    **for** $i = 0$; $i <$ iter; $i$++ **do**
5:       Perturb cur-pt to get new-pts /* Random swap and transformation (XX/180°, XY/180°) */
6:       Remove non-CC solutions from new-pts
7:       Remove solutions in new-pts having diffusion break and dummy count above upper bounds
8:       Keep only non-dominated solutions in new-pts
9:       // Assume size of new-pts is now $k$
10:      /* update cur-pt */
11:      **if** cur-pt dominates $k_1$ (> 0) solutions in new-pts **then** /* Case 1 */
12:         Randomly pick a solution new-pt from $k - k_1$ solutions
13:         Assign new-pt to cur-pt with probability=$prob$ (Eq. (2))
14:      **end if**
15:      **if** cur-pt and new-pts are non-dominating to each other **then** /* Case 2 */
16:         Randomly pick a solution new-pt from new-pts
17:         Assign new-pt to cur-pt with probability=$prob$ (Eq. (2) with $k_1 = 0$)
18:      **end if**
19:      **if** $k_1$ (> 0) solutions in new-pts dominate cur-pt **then** /* Case 3 */
20:         Randomly pick a solution new-pt from $k_1$ solutions
21:         Assign new-pt to cur-pt
22:      **end if**
23:      /* update archive */
24:      Replace the dominated solutions in archive with new-pts
25:    **end for**
26:    temp = $\alpha$ * temp
27: **end while**
28: Get the desired solution from the archive

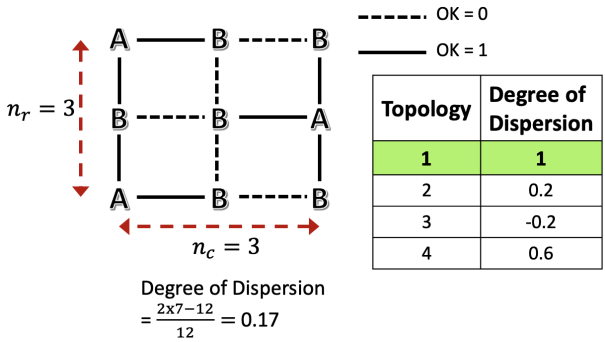

Figure 5: Defining degree of dispersion for devices A and B having 3 and 6 units respectively. Degree of Dispersion values for four layouts from Figure 3. Topology 1 has the highest degree of dispersion.

the transistor to the well-edge, as illustrated in Figure 6.

$$\Delta V_{th} \propto \frac{1}{\text{WPE}} = \sum_{i=1}^{n} (\frac{1}{SC_L^i + L_g} + \frac{1}{SC_R^i + L_g} + \frac{1}{SC_T^i + W_g} + \frac{1}{SC_B^i + W_g}) \quad (4)$$

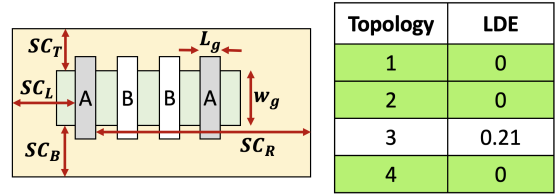

Figure 6: Well Proximity Effect (WPE). The Table shows Layout Dependent Effects (LDE) that capture Length of Diffusion (LOD) and Well Proximity Effect (WPE) for the four topologies from the Figure 3.

Here, $n$ represents the number of unit cells, while $L_g$ and $W_g$ denote the gate length and width of a unit, respectively. $SC_L$, $SC_R$, $SC_T$, and $SC_B$ are the distances from the left, right, top, and bottom well boundaries. To mitigate the threshold voltage variations ($\Delta V_{th}$) introduced by the WPE, we minimize the difference, or mismatch, of the mean values of $\frac{1}{\text{WPE}}$ across all devices, captured by the following model.

$$\sum_{k=1}^{N-1} \sum_{l=k+1}^{N} |\frac{(\frac{1}{\text{WPE}})_k}{n_k} - \frac{(\frac{1}{\text{WPE}})_l}{n_l}| \quad (5)$$

Here, $N$ is the number of devices and $n_i$ is the number of unit cells of the device $i$. $(\frac{1}{\text{WPE}})_i$ is the $\frac{1}{\text{WPE}}$ value of the device $i$. The following simplification of WPE is sufficient for comparison purpose.

$$(\frac{1}{\text{WPE}})_i = \sum_{u=1}^{n_i} (\frac{1}{x_u} + \frac{1}{r + 1 - x_u} + \frac{1}{y_u} + \frac{1}{c + 1 - y_u}) \quad (6)$$

$x_u$ and $y_u$ are respectively the column and row of the unit cell $u$, $r$ and $c$ are the number of unit cells in each row and column, respectively. Note that the first two terms also capture the Length of Diffusion [22][25]. For the four layouts in Figure 3 the LDE values are shown in the Table in Figure 6. Except topology 3, all other topologies have LDE value 0.

### 3.3 Routing Cost

In their work, [22] utilized the concept of rectilinear minimum spanning tree (RMST) to calculate routing cost. However, RMST can lead to overlapping edges, potentially costing up to 1.5 times longer wire length than minimum rectilinear steiner tree (MRST) [2]. Although calculating the steiner tree is an NP-complete problem, many fast heuristics have been proposed. We use one of the very first algorithm for steiner tree calculation proposed in [2] due to ease of implementation and suitability for the problem size of ours.

The algorithm begins by constructing an undirected graph, where each unit cell connected to the net is represented as a node. Rectilinear edges connect every pair of nodes on the net, and the RMST is obtained using Prim's algorithm. Subsequently, for each node ($n$) and edge pair ($u,v$) in the RMST, a new steiner node that improves the wirelength is obtained. The shortest path ($sp$) from the node ($n$) to the edge ($u,v$) or the rectangular layout of the edge is then calculated. The newly introduced node ($p$) on the edge or the rectangle becomes the steiner node. The edge $(u,v)$ is replaced by the edges $(p,u)$, $(p,v)$, and $(n,p)$, forming a cycle. The algorithm iteratively identifies and removes the edge with the largest weight $(u,n)$ until no further improvement is observed. For the example in Figure 7, removing $(u,n)$ from the cycle $n$->$p$->$u$->$n$ results in a gain of weight$(u,n)$ − weight$(n,p)$ = 4 − 2 = 2.

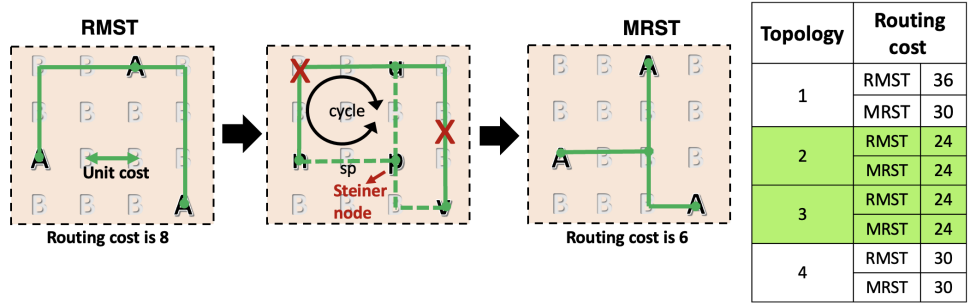

Figure 7: Routing cost calculation for device A having 3 units. Topologies 2 and 3 have the minimum routing cost. MRST yields smaller routing cost for the topology 1.

The routing costs for the four layouts from the Figure 3 are presented in Figure 7. Topologies 2 and 3 have the best routing cost due to more clustering of units A and B compared to other topologies. The trade-off between the routing cost and the degree of dispersion can be noted. MRST yields smaller routing costs than RMST for the topology 1. A pseudocode for the routing cost calculation is presented in **Algorithm 3**, which calls **Algorithm 2**.

---

**Algorithm 2 trialAddSteiner**

1: **trialAddSteiner** $(G, n, (u, v))$
2: find shortest distance node $p$ on rectangular layout of $(u, v)$ from $n$ /* $p$ is the steiner node */
3: $G.node \leftarrow p$
4: remove $(u, v)$ from $G$
5: $G.edge \leftarrow (p, u)$
6: $G.weight \leftarrow |loc.p.x - loc.u.x| + |loc.p.y - loc.u.y|$
7: $G.edge \leftarrow (p, v)$
8: $G.weight \leftarrow |loc.p.x - loc.v.x| + |loc.p.y - loc.v.y|$
9: $G.edge \leftarrow (n, p)$
10: $G.weight \leftarrow |loc.n.x - loc.p.x| + |loc.n.y - loc.p.y|$
11: find the edge with largest weight $(u, n)$ in $G$
12: remove $(u, n)$ from $G$
13: $gain$ = weight$(u, n)$ - weight$(n, p)$
14: **return** <$gain, G$>
15: **end trialAddSteiner**

---

## 4 RESULTS AND DISCUSSIONS

The proposed algorithm has been implemented in C++, and all experiments have been conducted on a Linux environment with Intel Core CPU running at 3.3 GHz with 128 GB of memory. In the **Algorithm 1**, we set $T_{max}$, $T_{min}$, $\alpha$, and $iter$ as 100, $10^{-7}$, 0.37 and 100, respectively. To benchmark our approach against state-of-the-art solutions [25], we have used five configurations of the current mirror structure (CM:1-5),

**Algorithm 3 Calculate Routing Cost**

Input : *topology, netlist*
1: $routing\_cost = 0$
2: **for** each *net* in *netlist* **do**
3: /* Construct the graph */
4: **for** each *unit* connected to *net* **do**
5: $G.node \leftarrow unit$
6: **end for**
7: **for** each pair of *node* $(u, v)$ in $G$ **do**
8: $G.edge \leftarrow (u, v)$
9: find location $(loc)$ of $u$ and $v$ in *topology*
10: $G.weight \leftarrow |loc.u.x - loc.v.x| + |loc.u.y - loc.v.y|$
11: **end for**
12: /* Find Rectilinear Minimum Spanning Tree (RMST) */
13: find $G_{RMST}$ in $G$ using Prim's algorithm
14: /* Find Minimum Rectilinear Steiner Tree (MRST) */
15: **while** there is improvement in sum$(G_{RMST}.weight)$ **do**
16: $i = 1$
17: **for** each $<n, (u, v)>$ in $G_{RMST}$ **do**
18: $<Gain[i++], tmp>$ = **trialAddSteiner** $(G_{RMST}, n, (u, v))$
19: **end for**
20: find max$(Gain)$ and corresponding $<n, (u, v)>$
21: $<tmp, G_{RMST}>$ = **trialAddSteiner** $(G_{RMST}, n, (u, v))$
22: **end while**
23: $G_{MRST} = G_{RMST}$
24: $routing\_cost = routing\_cost + $ sum$(G_{MRST}.weight)$
25: **end for**
26: **return** $routing\_cost$

---

from [25]. The results for the five test cases are shown in Table 1. Both algorithms generate solutions without any diffusion break. To validate the routing cost model, we use Cadence Virtuoso Layout Suite [5] to auto-place and auto-route the CC topology and measure the total routed wirelength. There is a good improvement in the routing cost, as predicted by both the routing model and the auto-router.

We report post-layout simulation results for the current mismatch using the industry-standard tool. We have used tsmc 40 nm technology for this purpose. For the current mirror circuit, we use mismatch expression $100 | n I_{src} - I_{dest} | / (n I_{src} + I_{dest})$, where $n$ is the current mirror ratio (e.g., for CM:1, $n = 11$), $I_{src}$ is the current flowing through the source transistor (e.g., $D_{T_0}$ in Figure 8(a)) and $I_{dest}$ is the total current flowing through the destination transistors (e.g., $D_{T_1}$ to $D_{T_n}$ in Figure 8(a)) which copy current from the source transistor. Ideally, the value should be 0 when the source and destination transistors match. With the exception of CM:1, in all test cases, there is an improvement in current mismatch value. It is important to note that the degree of dispersion, the layout dependent effects (LDE) and the parasitics all play a role in current matching performance. A smaller degree of dispersion suggests less uniform device spread, potentially degrading spatial variation performance. Parasitic resistance contributes to IR drop, thereby affecting transistor current, layout dependent effects affect transistor current by causing change in the threshold voltage. We do not consider random mismatch (such analysis is usually done by monte-carlo simulation) in our simulation, as CC layout is not useful for canceling random variation.

| Test case | Method | Degree of Dispersion | LDE | Mismatch in Current | Routing Cost Model | Routing Cost Router | Diffusion Break | Dummy Count | Layout Area |
|---|---|---|---|---|---|---|---|---|---|
| CM:1 [2,2,4,8,8], K=1.3 | [25] | 0.37 | 0.50 | 2.38 | 75 | 78 | 0 | 0 | 15 |
| | This work | 0.47 | 0.21 | 3.39 | 77 | 79 | 0 | 0 | 15 |
| CM:2 [2,2,4,10], K=2 | [25] | 0.04 | 0.59 | 3.1 | 55 | 63 | 0 | 0 | 12 |
| | This work | 0.19 | 0.46 | 0.7 | 55 | 63 | 0 | 0 | 12 |
| CM:3 [2,2,4,8], K=1.3 | [25] | 0.17 | 0.47 | 1.8 | 46 | 52 | 0 | 0 | 10 |
| | This work | 0.17 | 0.39 | 1.7 | 46 | 51 | 0 | 0 | 10 |
| CM:4 [4,4,8,8], K=1.3 | [25] | 0.58 | 0.58 | 3.72 | 76 | 74 | 0 | 0 | 14 |
| | This work | 0.26 | 0.34 | 0.1 | 69 | 71 | 0 | 0 | 14 |
| CM:5 [4,4,4,10,10], K=2 | [25] | 0.38 | 0.73 | 4.46 | 108 | 112 | 0 | 0 | 22 |
| | This work | 0.23 | 0.57 | 1.02 | 99 | 110 | 0 | 0 | 22 |

Table 1: Comparison with [25] for five current mirror (CM) configurations reported in [25].

| Test case | Method | Degree of Dispersion | LDE | Mismatch in Current | Routing Cost Model | Routing Cost Router | Diffusion Break | Dummy Count | Layout Area |
|---|---|---|---|---|---|---|---|---|---|
| CM:1 [2,2,2,2,10], K=2 | [25] | 0.19 | 0.97 | 2.04 | 56 | 78 | 2 | 6 | 15 |
| | This work | 0.33 | 0.92 | 1.06 | 53 | 73 | 2 | 6 | 15 |
| CM:2 [2,2,2,6,6], K=2 | [25] | 0.48 | 0.98 | 3.46 | 58 | 75 | 2 | 6 | 16 |
| | This work | 0.48 | 0.82 | 3.11 | 53 | 72 | 2 | 6 | 16 |
| CDIP:1 [6,6,10,10], K=2 | [25] | 0.54 | 0.21 | 1.05 | 124 | 186 | 4 | 8 | 28 |
| | This work | 0.69 | 0.01 | 0.02 | 122 | 139 | 0 | 0 | 24 |
| CDIP:2 [10,10,10,10], K=2 | [25] | 0.58 | 0.10 | 0.01 | 150 | 227 | 4 | 16 | 35 |
| | This work | 0.58 | 0.07 | 0.50 | 146 | 194 | 2 | 8 | 30 |
| CDLP:1 [2,2,6,6], K=2 | [25] | 0.33 | 0.56 | 0.10 | 47 | 66 | 0 | 0 | 12 |
| | This work | 0.17 | 0.56 | 4.49 | 43 | 57 | 0 | 0 | 12 |
| CDLP:2 [6,6,6,6], K=1.3 | [25] | 0.58 | 0.42 | 6.90 | 64 | 100 | 4 | 12 | 20 |
| | This work | 0.58 | 0.22 | 0.37 | 61 | 87 | 2 | 4 | 16 |

Table 2: Comparison with [25] for six random testcases: Current Mirror (CM), Cascode Differential Input Pair (CDIP), and Cascode Differential Load Pair (CDLP).

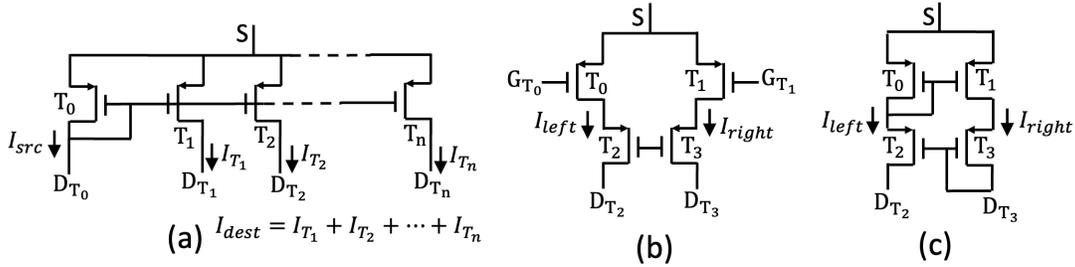

Figure 8: (a) Current Mirror (CM) (b) Cascode Differential Input Pair (CDIP) (c) Cascode Differential Load Pair (CDLP).

We have created an additional set of six tests, incorporating two scenarios from each of the three distinct configurations: Current Mirror (CM), Cascode Differential Input Pair (CDIP), and Cascode Differential Load Pair (CDLP) as shown in Figure 8. The results for these test cases are shown in Table 2. In half of the test cases, improvement in diffusion break or dummies is observed. This is particularly significant as a reduction in dummies means that the area usage is smaller, which can reduce routing cost, and the impact of variation. Note that our algorithm optimizes both the dummy count and the diffusion break. We have observed that having only dummy count as the optimization objective gives inferior results. This could be attributed to the diffusion break having a smaller search space, which helps the optimizer avoid getting stuck at local minima. Figure 9 shows how dummy count can increase when no constraint is set on the diffusion break. In our experiment, we have used the state-of-the-art solution as a constraint to limit the dummy count and the diffusion break. However, the initial placement solution can also be set as a constraint.

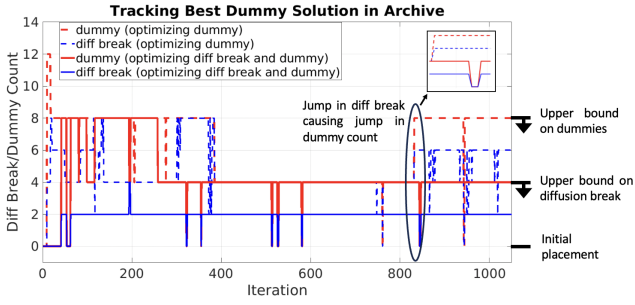

**Figure 9: Shows how limiting diffusion break to not go above the upper bound helps keep dummy count in check.**

Across different circuit configurations, there is a good improvement in routed wirelength. For measuring the current mismatch in CDIP and CDLP, we use the expression $100|I_{left} - I_{right}|/(I_{left} + I_{right})$, representing the difference in current between the left and right branches of the differential pair with the resultant difference normalized by the total current. Both Tables have more instances of green than red, signifying good improvement. It is important to underscore that routing cost, mismatch in current, and layout area serve as practical metrics for improvement. Across various circuit configurations, our algorithm performs better than state-of-the-art algorithm in these parameters. The runtime of our algorithm is about 1 min for the largest test case, CDIP:2.

### 4.1 Multi-Objective Optimization

Multi-objective optimization algorithm may produce numerous non-dominant solutions, necessitating the use of clustering techniques to reduce the number of solutions [1]. However, in the problem handled in this paper, we observe fewer than a hundred solutions, owing to the relatively smaller search space, especially considering integer nature of the diffusion break or dummy solution. We therefore do not employ the clustering technique. To obtain the optimized solution presented in Tables 1 and 2 from the set of solutions, we have done some post-processing, assigning greater weight to solutions having better diffusion break or dummies, routing cost, and Layout Dependent Effects. These selected solutions usually have worse degree of dispersion value, showing competing nature of the different performance metrics. There are solutions with a better degree of dispersion compared to state-of-the-art, however, such solutions have degradation in other critical parameters. Highlighting the distinction from single-objective optimization, where we indirectly influence the final solution by adjusting the objectives' coefficients, in the current approach, we have a set of solutions from which we select the best solution based on the target requirements. This is an important advantage over state-of-the-art methods, as they cannot produce competing or non-dominating solutions.

*4.1.1 Comparison to AMOSA ("one solution" approach).* Note that AMOSA [1] can be modified in a straightforward way to handle multiple solutions in each iteration of the optimization by randomly selecting one solution. However, we have observed that our approach to handling multiple solutions yields more and higher quality solutions, particularly in XY/180° transformation. For the probability calculation, the inverse effect of temperature in Eq. (2) in AMOSA [1] yields inferior results compared to Eq. (2).

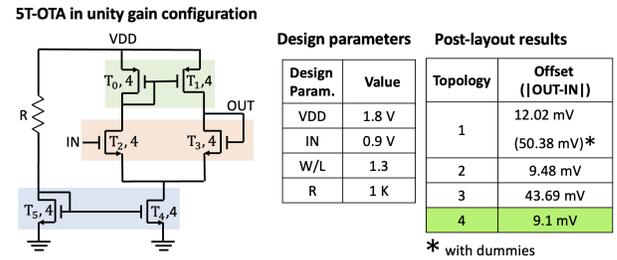

**Figure 10: A 5T-OTA shows best offset performance with CC topology 4.**

### 4.2 Post-layout Simulation Results of a 5T-OTA

We present post-layout simulation results of a 5T-OTA circuit in Figure 10. Each transistor in the OTA comprises four units with design parameters listed in the Table. They are arranged in groups (<$T_0$ $T_1$>, <$T_2$ $T_3$>, <$T_4$ $T_5$>), and laid out following the four CC topologies as illustrated before in Figure 3. The layout is done using the same Cadence Virtuoso auto-placer and auto-router [5]. To measure the offset voltage, the OTA is connected in unity gain configuration. A schematic level simulation of the OTA yields an offset voltage of 3.41mV. In post-layout, there is degradation in the offset as reported in the Table in Figure 10. Note that except topology 1, all other topologies can fully share diffusion region without using dummies. Topology 1 has 12.02 mV and 50.38 mV offset, respectively, without and with dummies. While topology 4 yields the best result, it is not always the ideal layout for all three groups in the OTA. The optimal result will likely come from employing different topologies for each group. This, however, is a circuit-level optimization problem where we have to also consider the impact of circuit block placement and routing on the final circuit performance—this is beyond the scope of the discussion here. Our point is that without exploring diverse topologies at device-level placement, achieving the best circuit performance is not possible. We have addressed the device-level placement issue in this paper.

# 5 CONCLUSIONS

We have introduced a unified multi-objective optimization framework for common-centroid placement of analog transistors. Our formulation includes important layout constraints. The proposed method enables exploration of more topologies by applying powerful transformations. In contrast to existing methods, our approach shows better performance in post-layout simulation, consistently generating more optimal CC placements across diverse circuit configurations.